\documentclass[aps,prl,twocolumn,a4paper,10pt]{revtex4-1}

\usepackage[T1]{fontenc}		%%% T1 Font Encoding
\usepackage[english]{babel}		%%% LaTeX for English
\usepackage[latin1]{inputenc}	%%% Accentuated Fonts
\usepackage{times}

\usepackage{amsmath}			%%% Mathematics AMS package
\usepackage{amssymb}			%%% Mathematics AMS package
\usepackage{bbm}				%%% Font for ensembes

\usepackage[colorlinks=true, a4paper=true, pdfstartview=FitV,linkcolor=blue, citecolor=blue, urlcolor=blue]{hyperref}

\usepackage{graphicx}			%%% Include graphics
\usepackage{graphics}			%%% Include graphics
\DeclareGraphicsExtensions{.pdf}
\usepackage{color}		

\newcommand{\bea}{\begin{eqnarray}}
\newcommand{\eea}{\end{eqnarray}}

\begin{document}
%%%%%%%%%%%%%%%%%%%%%%%%%%%%%%%%%%%%%%%%%%%%%%%%%%%%%%%%%%%%%%%%%%%%%%%%%%%%%%%%%%%%%%%%%%
%%%%%%%%%%%%%%%%%%%%%%%%%%%%%%%%%%%%%%%%%%%%%%%%%%%%%%%%%%%%%%%%%%%%%%%%%%%%%%%%%%%%%%%%%%
%%%% Title informations and authors
\title{Quantum Walk in Degenerate Spin Environments}

\author{Johan~Carlstr\"om,$^{1,2}$ Nikolay~Prokof'ev,$^{1,2,3}$ and Boris~Svistunov$^{1,3,4}$
}
\affiliation{
${}^1$ Department of Physics, University of Massachusetts, Amherst MA 01003 USA}
\affiliation{
${}^2$ Department of Theoretical Physics, The Royal Institute of Technology, Stockholm SE-10691 Sweden}
\affiliation{
${}^3$ National Research Center ``Kurchatov Institute,'' 123182 Moscow, Russia}
\affiliation{
${}^4$ Wilczek Quantum Center, Zhejiang University of Technology, Hangzhou 310014, China}
\date{\today}

\begin{abstract}
We study the propagation of a hole in degenerate (paramagnetic) spin environments.
This canonical problem has important connections to a number of physical systems, and is
perfectly suited for experimental realization with ultra-cold atoms in an optical lattice.
At the short-to-intermediate timescale that we can access using   a stochastic-series-type numeric scheme,
the propagation turns out to be distinctly non-diffusive with the probability
distribution featuring minima in both space and time due to quantum
interference, yet the motion is not ballistic, except at the beginning.
We discuss possible scenarios for long-term evolution that could be explored 
with an unprecedented degree of detail in experiments with single-atom resolved imaging.
 \end{abstract}
\maketitle

\begin{figure}[!htb]
\includegraphics[width=\linewidth]{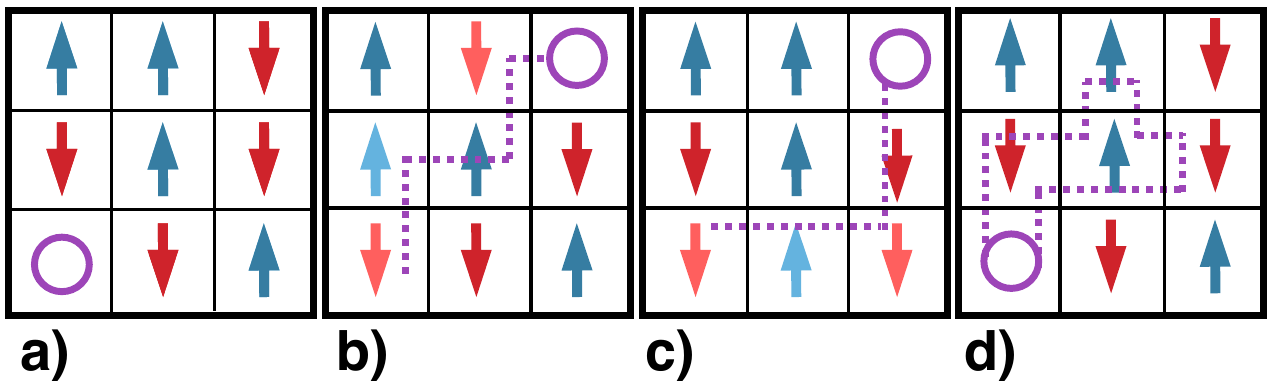}
\caption{
Hole propagation in a spin environment. (a) The hole starts at the lower left corner.
As it moves, it leaves behind a trace of altered spins. In (b,c), the final destination is the upper right corner, yet the paths taken differ (purple dashed lines), and so do
the resulting spin states. In (d), we show a self-retracing trajectory described as a
necklace of zero-area loops; all such trajectories interfere because at the end the hole
returns to the origin, and the spin environment is inherently preserved. 
}
\label{cartoon}
\end{figure}

%background
While classical random walks are well-understood as a diffusive process
realized in a wide range of physical systems, their quantum-mechanical counterparts
are far more subtle \cite{PhysRevA.48.1687}. To be more specific, consider
vacancy motion in the paramagnetic phase of solid $^3$He, or, equivalently,
hole propagation in the strongly correlated Mott-insulator (MI) state of electrons.
Even though the dynamics of holes/vacancies are governed by standard quantum
mechanics, the probability amplitudes for various trajectories do not interfere
the same way as they do for ballistic motion in a perfect (or spin polarized)
lattice because propagating holes often leave behind a trace in the spin environment that
effectively ``records'' where they have been (see Fig.~\ref{cartoon}).
This disrupts quantum interference between different paths, which, otherwise,
would lead to a much larger mean-square displacement than in the classical case
for the same path arc-length. However, some trajectories and even whole
classes of trajectories (see panel (d) in Fig.~\ref{cartoon}) do interfere.
This leads to a highly nontrivial propagation intermediate between the quantum-ballistic
and classical-diffusive limits.   
Interference between trajectories (leading to one and the same final state) 
may also depend on whether the lattice is bipartite or not, as well as on
the statistics of particles behind the spin components (see below).

In condensed matter systems that are too complicated/strongly interacting to allow
for exact solution, accurate analytic treatment, or viable numerical computations
in the thermodynamic limit, the study of hole propagation in model systems where
lattice sites carry an additional ``flavor'' index offers a means of gaining insight
into density of states, transport properties, formation of magnetic polarons, and
the nature of ferromagnetic instability in MI \cite{PhysRev.147.392}.
Characteristic examples include the Fermi Hubbard and $t-J$ models
\cite{PhysRev.147.392,brink},
the Kitaev-Heisenberg model \cite{PhysRevB.90.024404},
as well as vacancy motion in solid $^3$He crystals \cite{PhysRevLett.55.963,PhysRevB.35.3162,ref1}.
In quantum computing and information processing, relevant problems and algorithms
are also frequently formulated as quantum walks on a network  \cite{Farhi,Venegas-Andraca2012,PhysRevA.67.052307}.
Finally, hole propagation in a degenerate spin environment provides an important
physical realization of a system experiencing so-called dissipation-less
decoherence when the environment ``records'' particle trajectories with negligible
energy transfer \cite{0034-4885-63-4-204,PhysRevA.74.020102}.

%experiments with cold atoms
To the best of our knowledge, the problem of hole propagation in degenerate
magnetic environments is still far from being understood even at the conceptual
level despite a considerable long-standing interest in a variety of contexts.
On the one hand, such basic questions as the probability of return, dynamic
formation of magnetic domains, the value of the diffusion constant (if any)
and its dependence on the initial conditions, remain unanswered theoretically.
On the other hand, it is virtually impossible to obtain accurate
experimental information about hole dynamics and changes in its local spin environment
as it moves along for realistic condensed matter systems.
However, an experimental realization of relevant model systems in a controlled
setup with full access to all lattice sites is now possible using cold atoms/ions
trapped in an optical lattice and imaged with single-site resolution \cite{greiner0,greiner,bloch1,Karski174,PhysRevLett.103.090504,PhysRevLett.104.100503}.
Remarkably, the most interesting regime of a near-degenerate
spin environment also happens to be the least demanding in terms of
lattice parameters and system temperature, and is easily implementable with existing
technology. In fact, the fermionic MI phase corresponding to large $U/J$
(where $J$ is the hopping matrix element and $U$ is on-site repulsion) at temperature
$T > 4J^2/U$ (above the onset of strong anti-ferromagnetic correlations) was already
created some years ago \cite{esslinger1,bloch2}.

In this Letter, we address the topic of a quantum walk undertaken by the hole
in degenerate spin environments on a square lattice with the goal of establishing
precise data for hole dynamics over short-to-intermediate time-scales, testing
existing analytical predictions, discussing open questions and possible scenarios
concerning long-time dynamics, and motivating (apart from providing benchmarks)
future experimental studies.

Physically, our system corresponds to the $N$-component Hubbard model
\bea
H= -J\sum_{\langle i,j\rangle,\sigma} c^\dagger_{i,\sigma} c_{j,\sigma}+U \sum_i n_{i}^2 
\quad  (\sigma=1,2,\dots ,N), \quad 
\label{Hamiltonian}
\eea
deep into the MI phase, $U/J \gg1$, at high temperature $T \gg J^2/U$ (which is also
the parameter regime where direct single-site resolved imaging techniques work best).
All components are assumed
to have one and the same---either bosonic, or fermionic---statistics. For brevity we will refer to $\sigma$ as the spin index. Then, $c_{i,\sigma}^\dagger$ is the creation operator of a $\sigma$-particle on site $i$,
$n_{i \sigma}=c_{i,\sigma}^\dagger c_{i\sigma}$, and $\langle i,j\rangle$ stands for
pairs of n.n. sites. To obtain the time-dependent wave function $\psi(t)$,
we expand the evolution operator in the Taylor series:
\bea
U(t)=e^{-i H t}\, =\, \sum_n \, (-i)^n\, \frac{t^n }{n!} \, H^n \,.
\eea
In the $U/J\to \infty$ limit at unit filling factor when
doubly occupied sites are forbidden, the only allowed dynamic process
in (\ref{Hamiltonian}) is hole-propagation which can be viewed as a quantum walk that
alters the configuration of lattice spins. Then, on a square lattice
with coordination number $z=4$, one can describe $H^n$ as a sum
of $z^n$ possible trajectories. 
Using Monte Carlo simulation techniques, we sample all sums stochastically \cite{PSP:2060252} and classify trajectories according to their distinguishable final states.
This allows us to study
the evolution of the spatial distribution function for the hole over some
time interval, limited by the available memory resources. The specific
protocol is as follows: \\
 $\bullet$ We start by proposing $n$ from the Poisson distribution
   \bea
   p(n)=\frac{(zt)^n}{n!}\; e^{-zt}\, ,
   \eea
   where time is given in units of $1/J$; \\
 $\bullet$  Then, a random walk of the hole with $n$ steps is conducted
   (at each step the hole is moved randomly to one of the n.n. sites); \\
 $\bullet$  The final displacement of the hole, ${\mathbf r}$,
   and the resulting configuration of the spin environment, $\gamma$,
   are registered, and $(-i)^n$ is added to a bin corresponding to the
   $| {\mathbf r},\gamma \rangle$ state.  This contribution is real or imaginary depending on the parity of $n$;  \\
 $\bullet$  The procedure is repeated, and the set of generated states
   is used to construct the wave function
   \bea
   \psi(t)=\sum_{{\mathbf r},\gamma} A_\gamma ({\mathbf r},t) | {\mathbf r},\gamma \rangle \, ,
   \eea
   where amplitudes $A_{\gamma}({\mathbf r},t)$ are normalized to unity,
   $\sum_{{\mathbf r},\gamma} |A_{\gamma}({\mathbf r},t)|^2=1$.
   The spatial probability distribution for the hole position is then given by
   \bea
   \rho({\mathbf r},t)=\sum_{\gamma} |A_{\gamma}({\mathbf r},t)|^2 \; ;
   \label{probability}
   \eea
 $\bullet$  Finally, the entire procedure is repeated for multiple (256)
   randomly generated initial states $\vert {\mathbf r}=0, \gamma_{\rm in} \rangle$
   to obtain averaged results for $\rho({\mathbf r},t)$. 
   
    There is no extra sign associated with fermionic exchange cycles 
in our case because closed trajectories on bipartite lattice always result 
in an even number of exchanges. Moreover, since real and imaginary parts 
of $\psi (t)$ are based on trajectories with different parity of $n$, the probability distribution $\rho ({\mathbf r}, t)$ remains insensitive to particle statistics even if the lattice is not bipartite. 
However, if one considers hole propagation in a bose-fermi mixture, then the fermionic sign does matter. Also, if the initial state is a superposition 
$\psi(0)=\sum_{{\mathbf r}',\beta} C_{{\mathbf r}',\beta}  | {\mathbf r}',\beta \rangle $,
then particle statistics become important on non-bipartite lattices. On bipartite lattices
it is always possible to map between the ferminic and bosonic problems, even if the initial state is a superposition. 

%Results
%
\begin{figure}[!htb]
\includegraphics[width=\linewidth]{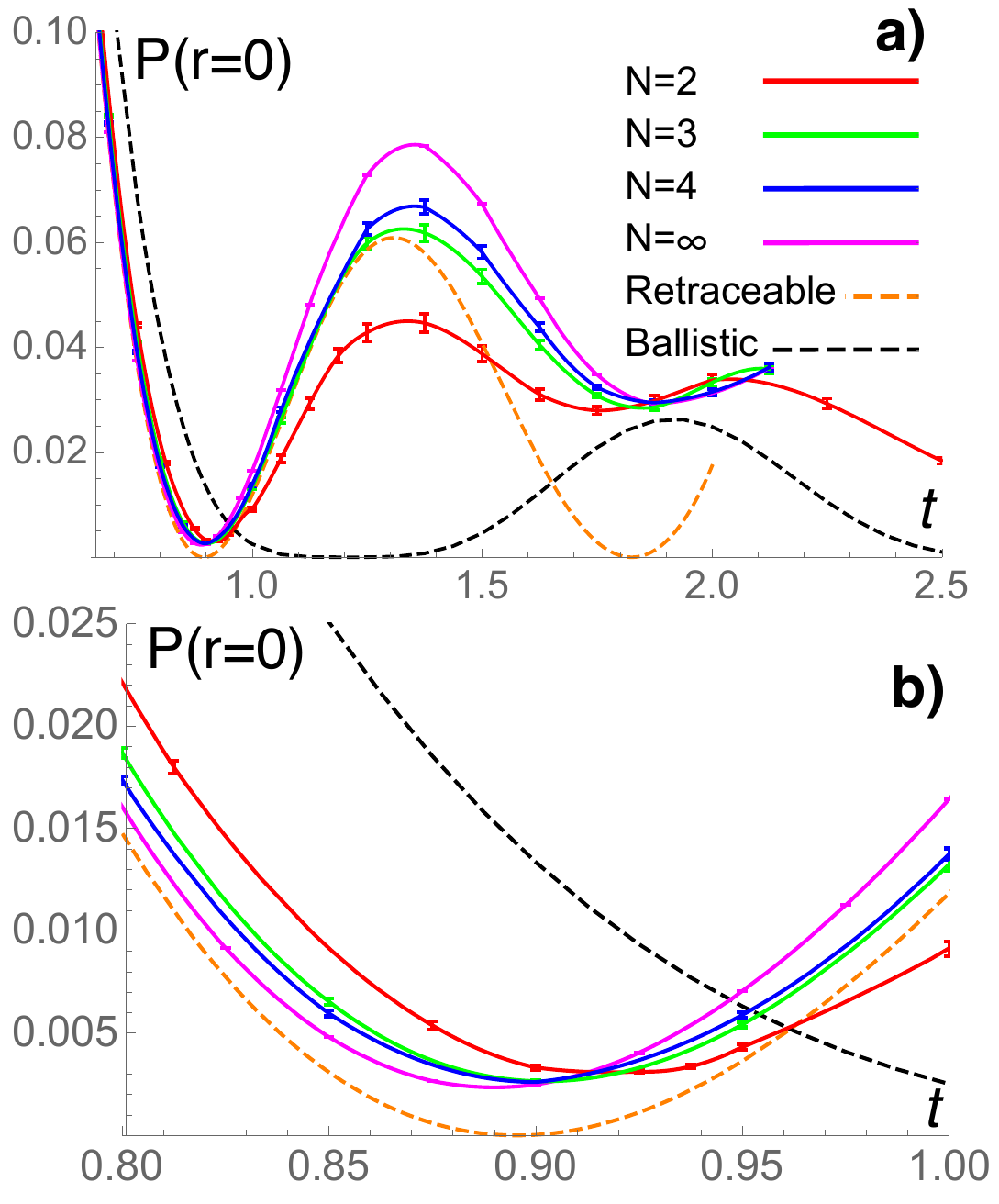}
\caption{
Probability of finding the hole at the origin as a function of time. The two panels  
show the same data for different time intervals.
The curves correspond to different spin values $N=2, \;3,\; 4\; \text{and } \infty$,
the BR approximation, and ballistic propagation (see legend). 
All curves, except that for ballistic motion, follow each other closely up to the first minimum at $t\approx 0.9$, but beyond that point, the details of the spin environment becomes important to the hole propagation. The BR approximation is very accurate up to the first minimum, but quickly fails at longer $t$.
}
\label{survive}
\end{figure}

In Fig.~\ref{survive}, we show the ``survival'' probability $\rho(0,t)$ for systems
with different number of spin components $N=2$, $3$, $4$ and $\infty$.
For comparison, we also include analytic results for ballistic propagation
in a perfect spin-polarized lattice as well as the Brinkman-Rice (BR)
approximation \cite{brink}. [The latter approximates
$\bar{A}(0,t)=\langle r=0,\gamma_{\rm in}\vert \psi(t)\rangle$
by considering only self-retracing closed paths
(see panel (d) in Fig.~\ref{cartoon})
that by construction preserve the original spin configuration $\gamma_{\rm in}$; then the spacial probability distribution is obtained through a second approximation given by
$\rho(0,t)\approx\vert \bar{A}(0,t) \vert^2$.]
Up to the first minimum, taking place at $t_{\rm min}\approx 0.9$,
the different scenarios are very similar, suggesting that the underlying spin degrees of freedom have little effect on the early evolution, because it is dominated by the
self-retracing paths that inherently preserve the spin environment.
Indeed, $n=0$, four $n=2$, and thirty two out of forty $n=4$ closed loop
trajectories are from the self-retracing set. Still, by excluding certain
trajectories from interference, the BR approximation predicts
$t_{\rm min}$ much more accurately than ballistic propagation, indicating
that dissipation-less decoherence starts playing a role at this time scale.
From Fig.~\ref{survive} (b) it is clear that all spin systems
exhibit near-complete destructive interference at the first minimum
with survival probabilities dropping below half a percent.

After the first minimum, the evolution is clearly dependent on the 
environment. Since the BR approximation only takes into account trajectories
that preserve the spin configuration, it goes completely wrong by predicting
instances of perfect destructive interference occurring at a later time
[an impossible effect due to the large number of $|\gamma \rangle$
states contributing to $\rho(0)$, see Eq.~(\ref{probability})].
The survival probability does show oscillations in time but they are strongly
damped, and, presumably, go away at longer time scales
(determining whether interference ceases to be relevant, and if
so than at what time scales, requires access to longer time scales).
Unexpectedly, the first oscillation is much weaker for $N=2$ than
for $N>2$. Intuitively, the probablity of finding the hole at the origin 
should be higher in systems with larger value of $N$ due to stronger 
decoherence effects; what comes as a surprise is the crossing of the
curves at $t\approx 1.9$.

\begin{figure}[!htb]
\includegraphics[width=\linewidth]{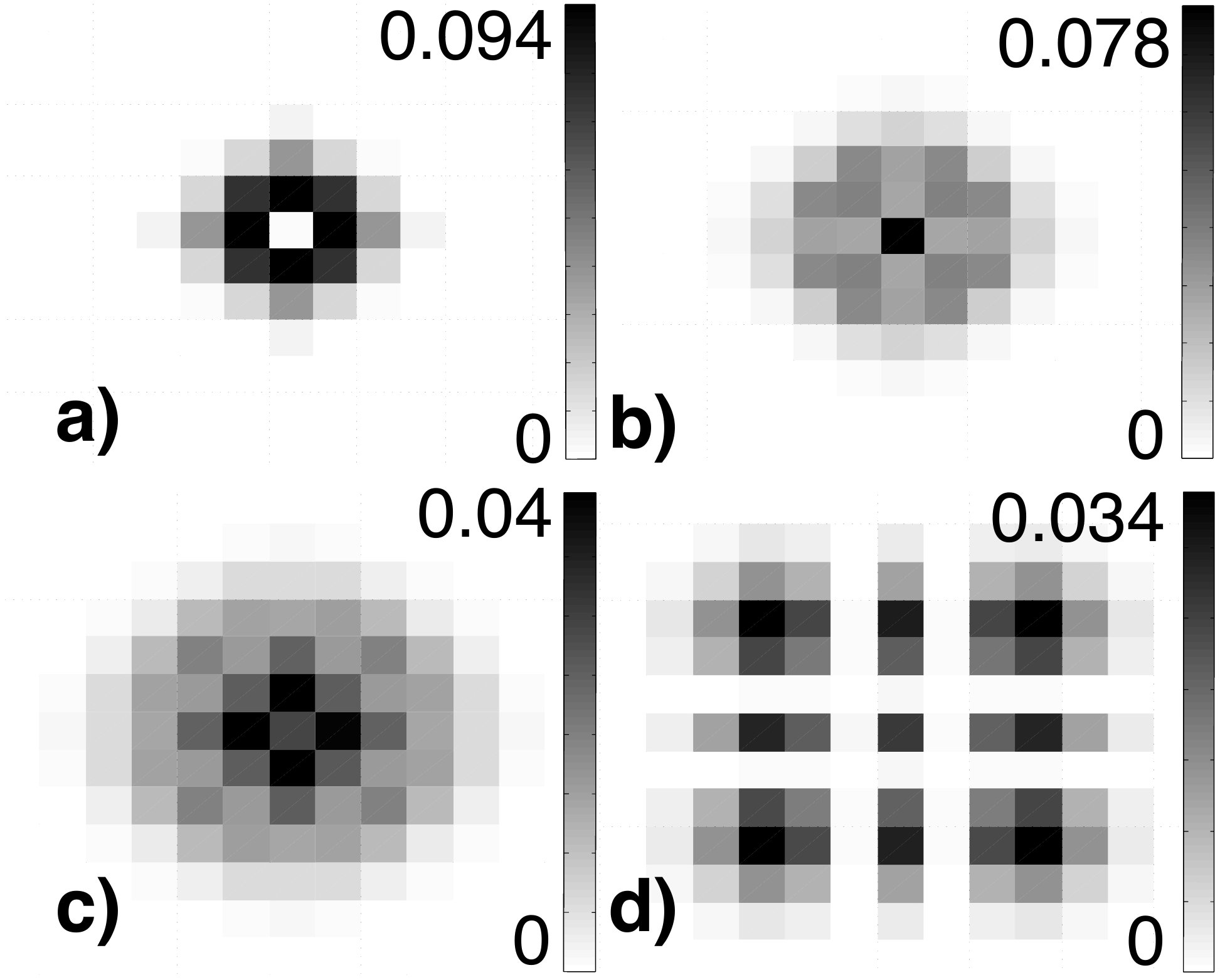}
\caption{
Spatial probability distributions $\rho(\mathbf{r},t)$ (Eq. \ref{probability}). 
Images (a-c) correspond to $N=\infty$ at $t=0.9,\; t=1.375$ and $t=1.875$, while (d) shows the case of ballistic motion in a polarised lattice at $t=2$. In (a), $\rho(0)$ is close to zero due to destructive interference, changing to maximum in (b) only to become a local minimum again in (c) at $t=1.875$, which also corresponds to a shallow local minimum in the survival probability,
see Fig.~\ref{survive}. In the ballistic case, strong destructive interference
is possible at arbitrary long times.
}
\label{wave}
\end{figure}
Minima and maxima in the survival probability are correlated with
minima and maxima in the spatial probability distributions, see Fig.~\ref{wave}.
The random walk in a degenerate spin environment is thus strikingly different
from the case of ballistic motion in a spin-polarized system that
features perfect destructive interference events at arbitrary long times (see Fig.~\ref{wave}d).
Yet the distribution function is not that of a classical random walk either,
which is just a gaussian without local minima in time or space.

\begin{figure}[!htb]
\includegraphics[width=\linewidth]{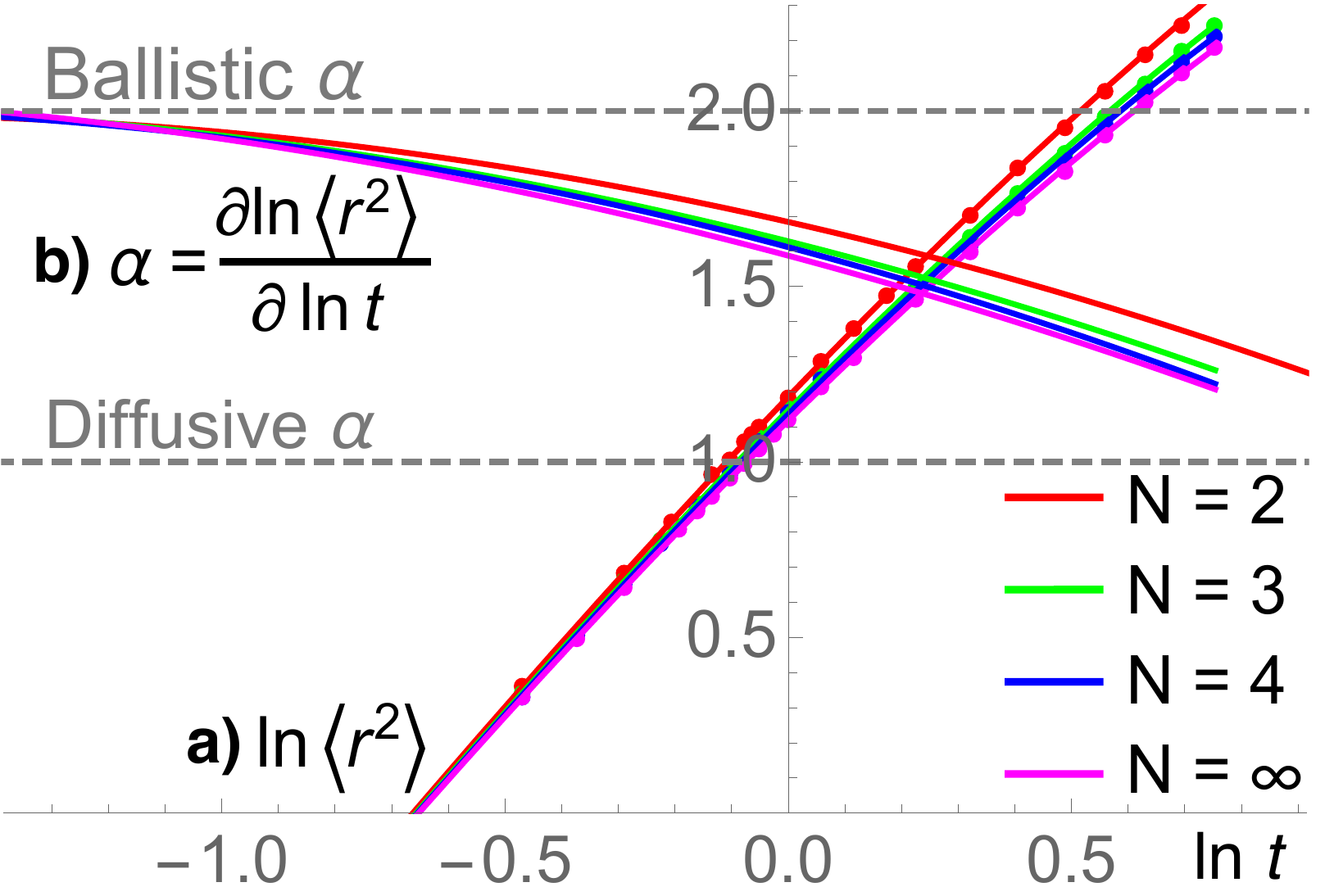}
\caption{
(a) Logarithmic plot of the mean-square displacement as a function of time.
The lines correspond to cubic polynomial fits to data.
(b) MSD exponent $\alpha$. At small $t$ it is close to $2$, as expected for coherent
propagation in a perfect lattice. 
Within the timescale of these simulations, $\alpha$ does not converge to a constant value, so the precise nature of the propagation at large $t$ (specifically if the motion is diffusive) remains an open question.  
}
\label{mobility}
\end{figure}
The most natural assumption is that in the long-time limit decoherence effects
ultimately lead to diffusive propagation characterized by linear dependence
of the mean-square displacement (MSD) on time. To see if our simulations have reached
this asymptotic behavior, we introduce the time-dependent MSD
exponent  $\alpha$ as the logarithmic derivative
\bea
\alpha = \frac{\partial \ln \langle r^2 \rangle}{\partial \ln t}\, .
\label{alpha}
\eea
Then, $\alpha$ can be deduced from the slope of the MSD curve on the log-log plot, see Fig.~\ref{mobility}a. Fitting a cubic polynomial to the data we can take the required
logarithmic derivative. The result is shown in Fig.~\ref{mobility}b.
Clearly, the diffusive regime was not yet reached within the timescale of our
simulations and we are still in the crossover region
(assuming that diffusion does take place, see below).
For $t\to 0$ the motion is nearly ballistic and $\alpha (t\to 0) \to 2$, as expected.

%impact on spin environment

Even though diffusion seems to be the most natural outcome of long time evolution,
this quantum walk has several features that make it fundamentally different from what
takes place in more conventional dissipative environments where external degrees of
freedom have dynamics of their own. In our case, {\it any} change in the spin environment is completely ``slaved'' to the hole dynamics. It is also different from coherent
propagation in static random media, because the spin environment does change under evolution.
The combination of these two circumstances leads to long-term memory effects
because previously created trajectory ``records'' can only be erased or scrambled by
the subsequent hole motion. These (and further) considerations make the problem of
long-time evolution in our system highly nontrivial.

\begin{figure}[!htb]
\includegraphics[width=0.4\linewidth, angle=-90]{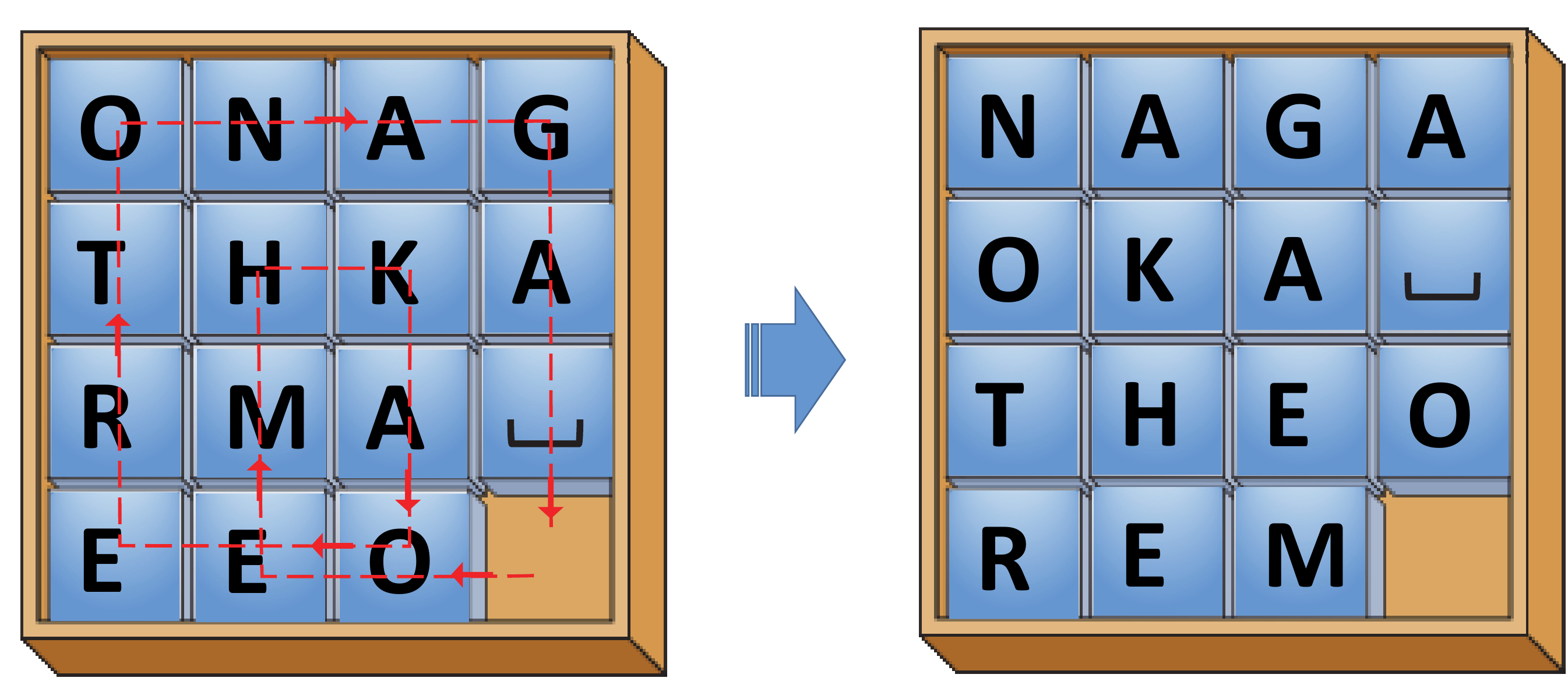}
\caption{
Hole propagation in a dynamically passive environment amounts to a quantum mechanical 
realisation of the N-puzzle game \cite{puzzle}, where a particular order is sought with as few tile movements as possible. While the classical problem is NP-hard, 
in the quantum mechanical system, all possible trajectories are realized in superposition.}
\label{npuzzle}
\end{figure}
Recall that in the $U/J \to \infty$ limit, the ground state of the Fermi-Hubbard model
with one hole added to the MI phase is a ferromagnet (Nagaoka theorem)\cite{PhysRev.147.392}
with a delocalized hole and a kinetic energy of $-zJ$.
Slightly higher energy states can be obtained by considering
ferromagnetic ``bubbles'' with delocalized holes inside.
These well-understood considerations immediately raise the question of whether
ferromagnetic ``bubble'' states play any role in the quantum walk problem considered here.
One way to see its relevance is to expand the initial state
$\psi(t=0) = \vert {\mathbf r}=0, \gamma_{\rm in} \rangle$  in terms of exact
eigenstates $|\epsilon \rangle$ and consider
$ \psi(t) = \sum_{\epsilon } e^{-i\epsilon t }C_{\epsilon} |\epsilon \rangle$.
Non-zero $C_{\epsilon}$ coefficients with $\epsilon \to -zJ$ are directly relating
random walk dynamics to ``bubble'' properties. Alternatively, one may consider
hole trajectories that assemble ferromagnetic ``bubbles'' from surrounding spins
and then spend most of the time inside them (it is easy to show that such trajectories
do exist for bubbles of arbitrary size and shape)---a kinematic Nagaoka effect. Note
also the close analogy between this phenomenon and the famous $n$-puzzle game, 
see Fig.~\ref{npuzzle} and \cite{puzzle}.
Even if this effect is unlikely to substantially alter the MSD
exponent, it may result in the bi-modal spatial probability distribution
and the survival probability decay slower than $1/t$ (corresponding to diffusion)
because ``bubbles'' are expected to diffuse much slower.

We did attempt to observe the quantum $n$-puzzle type physics in our simulations
by computing changes in the expectation value of the n.n. spin correlation
\bea
J=\sum_{\langle i,j\rangle} \sigma_i \sigma_j \, ,
\eea
and expecting to see its increase at long time due to the build up of ferromagnetic domains.
The average was further decomposed according to the final hole displacement to obtain
$\delta J ({\mathbf r},t) = \langle J \rangle ({\mathbf r},t) - \langle J \rangle(t=0)$ 
and check if
closed-loop trajectories are playing a special role.
By carrying out this computation, we did not detect a meaningful signal
within our error bars which are of the order $10^{-1}$, even at $r=0$ and $t=t_{\rm max}$.
We think that this outcome is mainly due to relatively short
trajectories explored in the simulations---too short for assembling ``bubbles''
with sizes significantly exceeding those happening for pure statistical reasons
within the initial state.

%conclusion

In conclusion, we find that at early stages the quantum walk of a hole in a degenerate spin environment is profoundly different from both the classical diffusive case and the quantum ballistic propagation in a polarized environment.
Whether the motion is diffusive in the long-time limit remains an open question.
We point out mechanisms that give rise to long-term memory effects
and strong correlations (interference) in the time domain, such as dynamic
assembly of ferromagnetic ``bubbles'' for some of the trajectories.
These effects, however, could not be addressed within the limited accuracy and
time scales of our simulations. Unitary evolution of the entire system further
complicates the observation of how holes ``solve'' the $n$-puzzle by quantum mechanics
because holes in random local spin configurations and holes in ``bubbles'' are
in a superposition state. One possible solution is to study dynamics of initial 
states with holes delocalized in ``bubbles''.  
 
We argue that real progress in understanding this long-standing problem is possible
by performing experiments with ultra-cold atoms in optical lattices
and studying the hole dynamics using single-site resolution imaging.
The most interesting parameter regime $(U/J \gg 1, T\gg J^2/U$,
deep in the MI phase at relatively high temperature) is already widely accessible.
In this limit, single-site imaging techniques work best and can provide an unprecedented amount of detail about hole-evolution as well as the state of the environment that is left behind.
Moreover, experimentally, one can reach time-scales orders of magnitude longer than
in our simulations, which can be used to benchmark experimental data at intermediate times.

We acknowledge support from the National Science Foundation under Grant PHY-1314735, the MURI Program ``New Quantum Phases of Matter''  from AFOSR
and the Wenner-Gren Foundation though a postdoctoral stipend.
Computations were performed on resources provided by the Swedish National Infrastructure
for Computing (SNIC) at the National Supercomputer Center in Link\"oping, Sweden.

\bibliography{biblio}

\end{document}